\documentclass[superscriptaddress,aps,9pt,twocolumn]{revtex4-2}

\usepackage[T1]{fontenc}
\usepackage{newtxtext}
\usepackage{newtxmath}
\usepackage{dsfont}
\usepackage{soul}

\usepackage{amsmath}
\usepackage{mathtools}
\usepackage{amsfonts}

\usepackage{qcircuit}
\usepackage{braket}

\usepackage{color}
\usepackage{xcolor}
\usepackage[colorlinks,citecolor=blue,linkcolor=blue,urlcolor=blue]{hyperref}

\usepackage{graphicx}
\usepackage[all]{hypcap} 

\usepackage[percent]{overpic}
\usepackage{empheq}
\usepackage[most]{tcolorbox}

\newcommand{\Tr}[1]{\text{Tr} #1}

\begin{document}

%\title{Work statistics of open quantum systems as a tool to }

\title{Measurement of the work statistics of an open quantum system using a quantum computer}

\author{Lindsay Bassman Oftelie}
\affiliation{NEST, Istituto Nanoscienze-CNR and Scuola Normale Superiore, 56127 Pisa, Italy}
\author{Michele Campisi}
\affiliation{NEST, Istituto Nanoscienze-CNR and Scuola Normale Superiore, 56127 Pisa, Italy}

\begin{abstract}
We report on the experimental measurement of the work statistics of a genuinely open quantum system using a quantum computer. Such measurement has remained elusive thus far due to the inherent difficulty in measuring the total energy change of a system-bath compound (which is the work) in the open quantum system scenario. We overcome this difficulty by applying the interferometric scheme, originally conceived for closed systems, to the open system case and implement it on a superconducting quantum computer, taking advantage of the relatively high levels of noise on current quantum hardware to realize an open quantum system. We demonstrate that the method can be used as a diagnostic tool to probe physical properties of the system-bath compound, such as its temperature and specific transition frequencies in its spectrum. Our experiments corroborate that the interferometric scheme is a promising tool to achieve the long-sought experimental validation of the Jarzynski equality for arbitrary open quantum systems.
\end{abstract}

\maketitle

\section{Introduction}
%A cornerstone of modern non-equilibrium thermodynamics (classical and quantum) is the established fact that non-equilibrium exact results such as the celebrated Jarzynski identity hold unaltered for open and closed systems, regardless of the strength of system-bath interaction \cite{Jarzynski04JSM04,Campisi09PRL102}. The reason is simple: the Jarzynski identity regards the statistics of work performed on a closed system. If your system is open, the total system-plus-bath system (shortly the SB system) can be considered as a closed system. Noting that the work (i.e. the energy injection from the external work source) is the change in energy in the SB system, it follows immediately that the Jarzynski identity holds when applied to the SB system \cite{Jarzynski04JSM04,Campisi09PRL102}. 

A cornerstone of modern non-equilibrium thermodynamics (classical and quantum) is that basic thermodynamic quantities, such as heat and work, are  stochastic (i.e., fluctuating) quantities. Their statistics are of crucial importance for the characterization of non-equilibrium phenomena at the microscopic level, as well as for understanding the microscopic origins of the second law of thermodynamics \cite{Jarzynski11ARCMP2,Campisi11RMP83}. %Notably, Jarzynski identity establishes that for any nonequilibrium process, the probability $p(w)$ to detect work $w$ contains the information required to compute equilibrium free energy differences. 
However, their measurement presents a plethora of difficulties and subtleties, depending on the specific experimental scenario. 

For instance, consider the work performed on a system subject to an external drive, in other words, the energy spent by an external agent to enact the drive  \cite{Jarzynski11ARCMP2,Campisi11RMP83}. In the case of a closed quantum system, the work can be measured by taking the difference between the energy of the system before and after the drive, according to the two-point projective measurement (TPM) scheme \cite{Campisi11RMP83}. This has been demonstrated experimentally both in a trapped ion system \cite{An15NATPHYS11}, as well as with clouds of cold atoms via an ingenious Stern-Gerlach type of set-up \cite{Roncaglia14PRL113,Cerisola17NATCOMMS8}, and can nowadays be readily performed with quantum computers \cite{Solfanelli21PRXQ2}. 

The situation gets considerably more complicated when the system is open, as in this case the work is equal to the total energy change of the compound system-plus-bath (hereon referred to as the SB system) across the non-equilibrium process (i.e., the drive). While it has been shown \cite{Jarzynski04JSM04,Campisi11RMP83} and experimentally demonstrated \cite{Liphardt02SCIENCE296,Collin05NAT437,Douarche05EPL70} that such measurements can be made in open classical systems by observing the dynamics of the driven classical system alone, measurements of the work in open quantum systems cannot be achieved in an analogous way \cite{Campisi11RMP83}. Instead, measuring the work generally requires two projective measurements of the total SB system energy \cite{Campisi09PRL102}. This is an extremely challenging task as one does not, in general, have access to (let alone control over) the very many degrees of freedom of the environment
\footnote{See however Refs. \cite{Hekking13PRL111,Suomela14PRB90} for an ingenious work-around in the case of a single qubit in the regime of weak coupling to the environment}. Therefore, experimental measurement of the work statistics in a non-equilibrium, open quantum systems has thus far remained elusive.

\begin{figure}\label{fig:schematics}
\centering
\begin{overpic}[width=0.93\columnwidth]{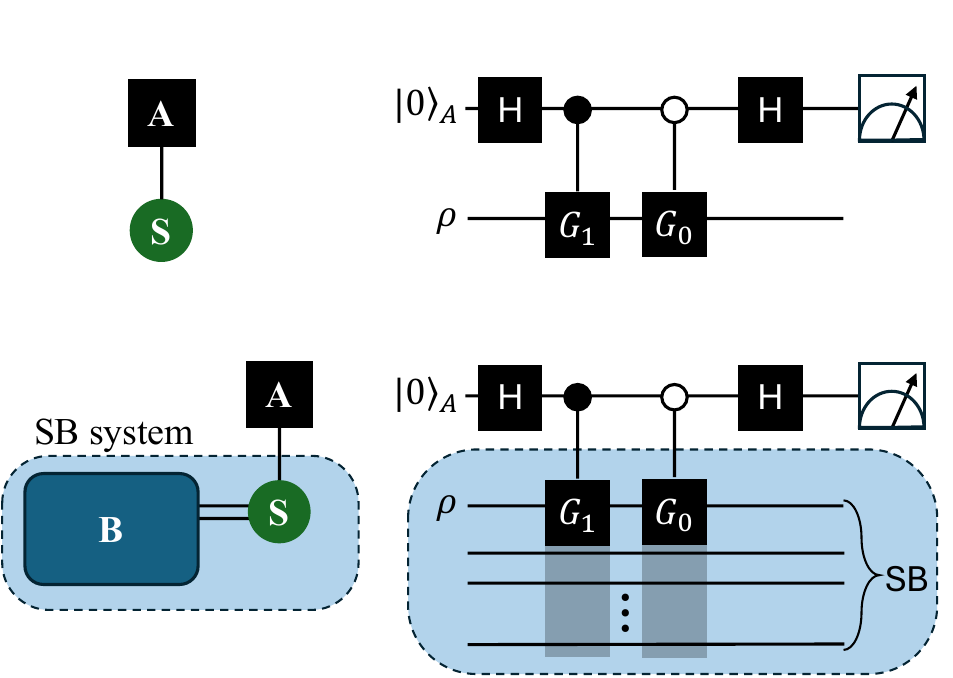}
\put(-1,70){(a)}
\put(37,70){(b)}
\put(-1,38){(c)}
\put(37,38){(d)}
\end{overpic}
\caption{Schematic diagram of the system and ancilla qubit employed in the interferometric scheme and the associated quantum circuits executing the scheme. (a) The ancilla qubit $A$ is connected to a closed system $S$. (b) Quantum circuit implementing the interferometric scheme to measure the work statistics of a closed quantum system $S$, realized with a single qubit. (c) The ancilla qubit $A$ is connected to an SB system comprising an open system $S$, which interacts with an environment $B$.  Crucially, the ancilla is not affected by the environment. (d) Quantum circuit implementing the interferometric scheme to measure the work statistics of an open quantum system, realized with a single qubit that interacts with its environment. }
\end{figure}
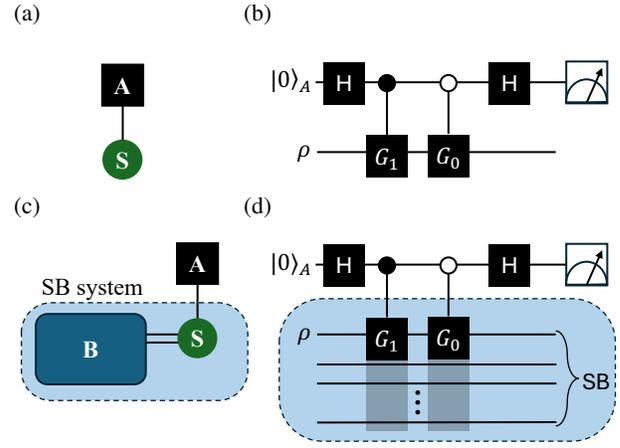

One possible strategy to overcome this difficulty is based on the so-called interferometric scheme, in which information about the work distribution is encoded into the state of an ancilla qubit via a properly designed interaction with the system of interest \cite{Dorner13PRL110,Mazzola13PRL110}. The method has been successfully applied in the case of a closed system  in an NMR experiment \cite{Batalhao14PRL113}.
Remarkably, as elucidated in Ref. \cite{Campisi15NJP17}, the ancilla is agnostic to the details of the system with which it is interacting; it simply encodes the total system energy change (i.e., work) statistics.  Thus, the system can be anything, from a single qubit, as shown schematically in Figure \ref{fig:schematics}a, to a generic composite system composed of a controllable system $S$ coupled to a bath $B$ (i.e., an SB system), shown schematically in Figure \ref{fig:schematics}c, as long as the ancilla $A$ is not itself affected by the presence of the bath. %As we shall detail further below, by engineering an experimental setup with a quantum SB system connected to an otherwise well isolated ancilla, the interferometric scheme can provide the long sought key to access the statistics of work of an open quantum system.   

Here, we report for the first time, measurements of the statistics of work in an open quantum system. Specifically, we measured the work statistics of an open qubit on a superconducting quantum computer via the interferometric scheme. Measuring the work statistics of an open quantum system is not a mere academic exercise: accessing it provides detailed information about the SB interactions, including the presence of specific transitions occurring within the SB system. %This information can in turn be used to improve the modelling of the open qubit dynamics. 
%It is important to remark that these cannot be captured by the reduced  dynamics of system $S$ alone, which only reflect the ``coarse grained'' effects of the SB transitions on $S$. Much of the information about global quantities, like work, which pertain the composite SB system gets lost when projecting onto the subsystem $S$ alone.
Indeed, our measurements reveal the presence of such SB transitions at the frequency of the qubits on the chip, suggesting that the open qubit is inadvertently interacting with other qubits, i.e., it is affected by cross-talk error.
Accordingly, the work statistics can be used as a diagnostic tool for the characterization of noise in quantum computers.

Furthermore, accessing the work statistics of open quantum systems is the key to unlock the experimental validation of the quantum Jarzynski equality for open systems \cite{Campisi09PRL102}, which remains an open challenge. Our results demonstrate that the interferometric scheme is a viable approach for reaching such a long-sought goal, and therefore constitute a considerable leap towards its achievement.
 
\section{theory}

\subsection{Closed case}
Consider a system, initially in state $\rho$, subject to a time dependent cyclical driving, such that its Hamiltonian reads
\begin{align}
    H(t)=H_0 + V(t)
\end{align}
where $V(t)=0$ for $t \notin [0,\tau]$.
The probability distribution function (PDF) of work $p(w)$ for such a system reads \cite{Campisi11RMP83}:
\begin{align}
    p(w)= \sum_{m n} p_{mn} \delta(w-E_m+E_n)
    \label{eq:pw}
\end{align}
where $p_{mn}$ is the joint probability of finding the eigenvalue $E_n$ in the projective measurement of $H_0$ at time $t=0$ \emph{and} the eigenvalue $E_m$ in the projective measurement of $H_0$ at time $t=\tau$ (i.e, the TPM scheme). In the following, for simplicity, we shall assume all eigenvalues to be non-degenerate. Consequently, we can write $p_{mn}= p_n p_{m|n}$ where $p_n= \langle n| \rho |n\rangle$ is the probability of obtaining $E_n$ in the first measurement, and $p_{m|n}= |\langle m| U  |n\rangle|^2$ is the conditional probability of finding the eigenvalue $E_m$ in time $\tau$, given that $E_n$ was observed at time $t=0$.

The interferometric scheme is a method to reconstruct the TPM work PDF, by encoding the necessary information into the state of an ancilla qubit. The encoding procedure, in the form of a quantum circuit, is illustrated in Fig. \ref{fig:schematics}b. Here, the system of interest $S$ is initially in the state $\rho$, and the ancilla qubit is prepared in the logical state $|0\rangle \langle 0|$ namely the eigenstate of the $\sigma_z$ Pauli matrix belonging to the eigenvalue $+1$. 
%The method consists of the application of the following gates: i) a Hadamard gate on the ancilla, ii) the conditional evolution $G_1= e^{-iuH_0} U \otimes |0\rangle \langle 0| + 1 \otimes |1\rangle \langle 1| $, iii) the conditional evolution $G_2= 1 \otimes |0\rangle \langle 0| + U e^{-iuH_0} \otimes |1\rangle \langle 1|$, and iv) a final Hadamard gate on the ancilla.
The method consists of the application of the following gates: i) a Hadamard gate on the ancilla, ii) the conditional evolution $G_1= \mathds{1} \otimes |0\rangle \langle 0| + e^{-iuH_0} U \otimes |1\rangle \langle 1|$, which applies the operator $e^{-iuH_0} U$ to the system qubit if the ancilla qubit is in the `1' state, iii) the conditional evolution $G_0= U e^{-iuH_0} \otimes |0\rangle \langle 0| + \mathds{1} \otimes |1\rangle \langle 1|$, which applies the operator $U e^{-iuH_0}$ to the system qubit if the ancilla qubit is in the `0' state, and iv) a final Hadamard gate on the ancilla. This quantum circuit encodes information about the quantity
\begin{align}
    g(u) = \Tr U^\dagger e^{iu H_0} U e^{-iuH_0} \rho
\end{align}
in the output ancilla state. Specifically, the expectation value of the ancilla Pauli $\sigma_z^A$ operator equals the real part of $g(u)$ while the expectation value of the ancilla Pauli $\sigma_y^A$ operator equals the imaginary part of $g(u)$ \cite{Dorner13PRL110,Mazzola13PRL110}:
\begin{align}
    g(u) = \langle \sigma_z^A \rangle(u) + i \langle \sigma_y^A \rangle(u)
    \label{eq:g=sigmas}
\end{align}
where we have made the dependence on the delay time $u$ of the averages explicit.

The quantity $g(u)$ can be re-written as:
\begin{align}
    g(u)= \sum_{m,n} e^{iu(E_m-E_n)} q_{mn}\\
    q_{mn} \doteq\langle m| U  |n\rangle \langle n| \rho U^\dagger  |m\rangle\, ,
\end{align}
which is the Fourier transform of the quantity
\begin{align}
    q(w)= \sum_{m n} q_{mn} \delta(w-E_m+E_n)\, .
    \label{eq:qw}
\end{align}
When $\rho$ commutes with $H_0$, that is when $\rho$ is diagonal in the $H_0$ eigenbasis, we have
$
   q_{mn}=\langle m| U  |n\rangle \langle n| \rho U^\dagger  |m\rangle = p_n p_{m|n}=p_{mn}
$
and accordingly $q(w)$ boils down to the TPM work PDF $p(w)$, Eq. (\ref{eq:pw}). Similarly, $g(u)$ boils down, in that case, to the work characteristic function $\chi(u)$, namely the Fourier transform of the TPM work PDF:
\begin{align}
    \chi(u) = \int_{-\infty}^{+\infty} dw e^{iuw} p(w) = \sum_{m,n} e^{iu(E_m-E_n)} p_{mn}\, .
\end{align}

Thus, provided the initial state $\rho$ commutes with $H_0$, one can access the characteristic function of work $\chi(u)$ at point $u$ with the interferometric scheme. By sampling the characteristic function for a sufficiently large and dense range of distinct values of $u$, one can then, by inverse Fourier transform, obtain an estimate of the TPM work PDF $p(w)$. 

Note that in general, the quantity $q_{mn}$ is not a joint probability. In fact, $q_{mn}$ is generally a complex number. However, its marginals $q''_m = \sum_n q_{mn}$ and $q'_n= \sum_m q_{mn}$ are proper discrete probability distributions, and is thus often referred to as a quasi-probability. Accordingly, the quantity $q(w)$ does not, in general, represent a proper PDF, nor is $g(u)$ a proper characteristic function.  We shall call them the ``quasi work PDF'' and the ``quasi work characteristic function'', respectively.

%The advantage of this procedure is that does not involve projective measurements on the system, but only projective measurements on the ancillary qubit at the end of the circuit. This is advantageous for two reasons: 1) the initial projective measurement could be so invasive as to impede the continuation of the protocol, 2) measuring the system could be extremely difficult if not impossible when the system itself is too large and uncontrollable, e.g. when it is a SB system, which is the case of interest for the present work.

In order to see how the method applies to the open quantum system case, consider, without loss of generality, the prototypical case when $V(t)=\lambda(t)Q$ (with $\lambda$ a real function such that $\lambda(t)=0$ for $t \notin (0,\tau)$ and $Q$ a system operator). As thoroughly illustrated in Ref. \cite{Dorner13PRL110}, in the closed case, the interferometric scheme is implemented by the evolution generated by the following System-Ancilla Hamiltonian in the time interval $t \in (0,\tau+u)$
\begin{align}
    H_{SA}(t) = H_0 \otimes \mathds{1}_A + f_1(t) Q \otimes \Pi_1 + f_0(t) Q\otimes \Pi_0\, ,
    \label{eq:HSA}
\end{align}
where $\Pi_{0}, \Pi_{1}$ are the projectors onto the ancilla states $|0\rangle$, $|1\rangle$, respectively, and 

\begin{align}
\begin{split}
  f_1(t) = 
    \begin{cases}
      \lambda(t), & 0 \le t < \tau \\
      0, & \tau \le t \le \tau + u\\ 
    \end{cases} 
    \\
    f_0(t) =
    \begin{cases}
      0, & 0 \le t < u \\
      \lambda(t-u), & u \le t \le \tau + u\\ 
    \end{cases}  
\end{split}
\end{align} 

Basically, the method consists in running the protocol $\lambda(t)$ for a time $\tau$ followed by a delay time $u$ during which $\lambda$ is kept at zero, conditioned on the ancilla  being in state $0$; and keeping $\lambda$ at zero for a time $u$ followed by the application of $\lambda(t)$ for a time $\tau$, conditioned on the ancilla being in state $1$. Thus $u$ is, physically, the delay time between the application of the driving $\lambda(t)$ in the two subspaces of the ancilla, which is responsible of the interference pattern collected in the state of the ancilla itself. Accordingly $\tau+u$ is the total time duration of the interferometric circuit. As shown in Refs. \cite{Dorner13PRL110, Mazzola13PRL110}, the Hamiltonian (\ref{eq:HSA}) imprints in the ancilla state, the function $g(u)$ associated with the driven closed quantum system $H_S(t)=H_0+\lambda Q$.

\subsection{Open case}

If you apply the very same driving described above but your system S is open (i.e., interacting with a bath), then the Hamiltonian that properly describes the $SBA$ dynamics (SB system plus ancilla), assuming the ancilla does not interact with the bath, is
\begin{align}
    H_{SBA}(t) &=\label{eq:HSBA}\\
    &H_0^{SB} \otimes \mathds{1}_A + f_1(t) Q \otimes \mathds{1}_B \otimes \Pi_1 + f_0(t) Q \otimes \mathds{1}_B \otimes \Pi_0 \nonumber
\end{align}
where%$Q^{SB}= Q \otimes \mathds{1}_B$, and 
\begin{align}
   H_0^{SB}=H_0^S \otimes \mathds{1}_B +\mathds{1}_S \otimes H_B  +H^{SB}_I \, ,
   \label{eq:H0SB}
\end{align}
with $H_B$ the bath Hamiltonian and $H^{SB}_I$ the SB interaction Hamiltonian.

Upon inspection of Eq. (\ref{eq:HSBA}), we recognize fact (and this is the most crucial point of our whole analysis) that such a Hamiltonian generates a quantum circuit (see Fig. \ref{fig:schematics}d) that encodes in the ancilla state, via Eq. (\ref{eq:g=sigmas}), the function $g(u)$ associated to the the driven open quantum system
\begin{align}
    H_{SB}(t) = H_0^{SB} + \lambda(t) Q \otimes \mathds{1}_B\, .
\end{align}

In other words, by enacting one and the same driving on the $SA$ compound, namely the $SA$ driving 
$f_1(t) Q \otimes \Pi_1 + f_0(t) Q\otimes \Pi_0 $ , one gets \emph{for free} either the closed-S system work PDF or the open-S system work PDF, depending on whether the S-system is closed or open, respectively %. This fortunate, yet subtle fact, was first noted in Ref. 
 \cite{Campisi13NJP15}.
More precisely, if the initial SB preparation commutes with $H_0^{SB}$, then $g(u)$ is indeed the work characteristic function $\chi(u)$ of the driven open system, and its Fourier transform gives the corresponding TPM work PDF, accounting for the energy changes $w=E_m^{SB}-E_n^{SB}$  occurring in the whole SB system ($E_m^{SB}$ denoting the eigenvalues of $H^0_{SB}$). 

\subsection{Dealing with the constraint $u\geq 0$}

We remark that according to the implementations of Eqs. (\ref{eq:HSA}) and (\ref{eq:HSBA}) for closed system $S$ or open system SB, respectively, $g(u)$ can only be sampled only for positive values of $u$, as $u$ represents a positive delay time. However, in order to calculate the inverse Fourier transform of $g(u)$ one needs to sample $g$ over the entire real axis.
%\begin{align}
%    q(w) = \frac{1}{2\pi} \int_{-\infty}^{+\infty} du e^{-iuw} g(u)
%\end{align}
This is not a problem because sampling over the positive real axis suffices to recover the quasi-probabilities $q_{mn}$, and hence the quasi-PDF $q(w)$. To see this, consider the half-inverse-Fourier transform of $g(u)$:
\begin{align}
    q_>(w) &= \frac{1}{2\pi} \int_0^\infty du  e^{-iuw} g(u) \label{eq:q_>} \\
        & =\frac{1}{2\pi} \sum_{mn} q_{mn} \int_0^\infty du  e^{-iu(w-E_m+E_n)}\, .
\end{align}
Using the well known formula:
\begin{align}
    \mathcal{F}^{-1}[\theta(u)](x) = \frac{1}{2\pi} [\pi \delta(x) + i \text{P} \frac{1}{x}]\, ,
\end{align}
where $\theta$ is the Heaviside step function and P denotes Cauchy's principal value, we get:
\begin{align}
    q_>(w) %&= \frac{1}{2\pi} \int_0^\infty du  e^{-iuw} g(u)\\
        & =\frac{1}{2\pi} \sum_{mn} q_{mn} \left[\pi \delta(w-E_m+E_n) + i \text{P}\frac{1}{w-E_m+E_n}\right].
\end{align}
Taking the real part we obtain:
\begin{align}
    &\mathbb{RE}[q_>(w)]= \label{eq:re-q_>}\\
    &\frac{1}{2\pi} \sum_{mn}
    \left[\pi \mathbb{RE}[q_{mn}]  \delta(w-E_m+E_n) -\mathbb{IM}[q_{mn}]  \text{P}\frac{1}{w-E_m+E_n}\right] 
%    &\mathbb{IM}[q_>(w)]= \\
%    & \frac{1}{2\pi} \sum_{mn}
%   \left[\pi \mathbb{IM}[q_{mn}]  \delta(w-E_m+E_n) + \mathbb{RE}[q_{mn}]  \text{P}\frac{1}{w-E_m+E_n}\right] \nonumber
\end{align}
from which one can infer the real and imaginary parts of $q_{mn}$, thereby fully reconstructing the quasi work PDF $q(w)$, Eq. (\ref{eq:qw}). Note that when the imaginary part of $q_{mn}$ is null, Eq. (\ref{eq:re-q_>}) reduces to
\begin{align}
    2\mathbb{RE}[q_>(w)]=p(w), \quad \text{for}\quad \mathbb{IM}[q_{mn}]=0.
    \label{eq:REq=p}
\end{align}
This occurs precisely when the initial state of the system does not contain any coherence.  Thus, in the absence of any initial coherence, the real part of the half-inverse-Fourier transform directly gives the work PDF.  A non-zero imaginary part of $q_{mn}$ can be identified in a plot of $\mathbb{RE}[q_>(w)]$ by the presence of antisymmetric $1/x$-type broadening around the peaks.  When the peaks of $\mathbb{RE}[q_>(w)]$ are instead symmetric, it is a clear signal that the $q_{mn}$ are all real, implying that there was no coherence in the initial preparation, and thus Eq. (\ref{eq:REq=p}) holds.  
%WE DONT NEED TO SAY THISTo summarize, when applied to an open system, the interferometric scheme either provides the TPM work PDF or instead, signals the presence of coherence in the initial state relative to the bare SB Hamiltonian $H_{0}^{SB}$. In either case, useful information can be gleaned about the interactions of the system with its environment.

\section{results}

We demonstrate the utility of the interferometric scheme in measuring the work PDF of an open quantum system with the IBM quantum computer. Ideally, the qubits of a noise-free quantum processor are identical (but distinguishable) two-level systems with infinite decoherence and relaxation times, which can be coupled to their nearest neighbors in a controllable, time-dependent manner but are otherwise completely isolated from all other qubits and, importantly, from the environment.  Any selected subset of the qubits of the processor can be viewed as a closed system that evolves unitarily, controlled by the computational program encoded in the quantum circuit that is executed.  In reality, quantum processors are noisy, open systems that interact with their environment, leading to finite decoherence and relaxation times of the qubits. Worse, spurious interactions may occur between qubits, known as cross-talk, which is currently a leading source of error in superconducting quantum computers \cite{zhao2022quantum, ketterer2023characterizing}.  

The qubits on the IBM quantum processors are continually calibrated every few hours to assess the levels of noise they experience, providing estimates for their decoherence and relaxation times, readout error probabilities, and fidelities of native gates enacted upon them.  There can be large variations in these values across the individual qubits, sometimes up to an order of magnitude. To a good approximation, qubits with large decoherence times (i.e., much larger than the total wall-clock time of the quantum circuit) can be considered ideal qubits that are disconnected from their environment, while qubits with small decoherence times (i.e., on the order of the total wall-clock time of the quantum circuit), for whatever reason have a stronger interaction with their environment, and can therefore be considered open quantum systems.  We are therefore able to demonstrate the measurement of the work PDF of an open quantum system using the interferometric scheme by carefully selecting a pair of neighboring qubits, one "open" qubit to serve as the open system and one "closed" qubit to serve as the ancilla, based on current calibration of their decoherence times.  

To perform our experiment, we select a pair of neighboring qubits on the \textit{ibm$\_$sherbrooke} quantum processor consisting of an ancilla qubit measured to have a decoherence time of 372 $\mu$s, and an open system qubit measured to have a much smaller decoherence time of 38 $\mu$s.  To measure the work PDF using the interferometric scheme we prepare a set of circuits of the form illustrated in Figure \ref{fig:schematics}b, sampling a range of different values for the time delay $u$ and using the driving protocol
\begin{align}
    U = \sqrt{\sigma_x} = \frac{1}{2}\begin{bmatrix}
                            1+i & 1-i \\
                            1-i & 1+1 
\end{bmatrix}
\end{align}
to construct the controlled $G_1$ and $G_0$ gates.  Crucially, if the $S$ qubit is, in fact, an open system, this will result in the physical execution of the circuit depicted in Figure \ref{fig:schematics}d, as predicted by the theory discussed above, albeit with the environmental and interaction Hamiltonians ($H_B, H_{SB}$) remaining unknown.
As will become clear below, this is indeed the case.  

The system and ancilla qubits are prepared in the their ground states, and the entire set of circuits is executed twice on the pair, using 1024 shots per circuit.  In the first execution, the $\sigma_z$ operator is measured on the ancilla qubit for each value of $u$. Averaging over the measurements for each value of $u$ provides an estimate of the real part of the quasi work characteristic function $g(u)$ of the driven open quantum system.  This was immediately followed by the second execution of the set of circuits, which now measure the $\sigma_y$ operator of the ancilla for each value of $u$, thus providing an estimate of the imaginary part of the quasi work characteristic function.  Taking the half-inverse-Fourier transform of the measured $g(u)$ as in Eq. (\ref{eq:q_>}), gives the quasi work PDF $q(w)$ as described above.  As noted in Eq. (\ref{eq:REq=p}), this will provide the true work PDF $p(w)$ if there is no coherence in the initial state of the system.

In general, in the ideal (i.e., noise-free) case in which the qubit is a closed system, the work PDF can have at most three peaks, centered around $w=0$, $w=\hbar \omega$, and $w=-\hbar \omega$:
\begin{align}
    p^{id}(w)&= (p_{0|0}^{id} p^{id}_0+p_{1|1}^{id} p^{id}_1) \delta(w)\nonumber \\
    &+  p_{0|1}^{id}p^{id}_1 \delta(w+\hbar \omega )+ p_{1|0}^{id}p^{id}_0 \delta(w-\hbar \omega ) \label{eq:p-3peaks}
\end{align}
where $\omega$ is the resonant frequency of the qubit, $p^{id}_a$ is the probability that the ideal qubit is initialized in the $a$ state, and $p_{b|a}^{id}$ is the probability that the ideal qubit is measured in the $b$ state after the drive, given that it was initialized in the $a$ state (where $a,b \in [0,1]$).

With our choice of $U$, $p_{b|a}^{id}=1/2$ for all choices of $a,b$.  Furthermore, if we assume the qubit is initialized in the `0' state, then we have $p^{id}_0=1$ and $p^{id}_1=0$.  Given these conditions, the resulting work PDF for the ideal closed system will only contain two peaks of intensity $1/2$, located at $w=0$ and $w=\hbar \omega$:
\begin{align}
    p^{id}(w)&= (1/2)\delta(w) + (1/2) \delta(w-\hbar \omega ).
    \label{eq:p-2peaks}
\end{align}
In our experiments, the resonant frequency of the qubit is $f = 4.85$ GHz (where $f = \omega/2\pi$), which sets $hf = \hbar\omega = 20.04$ $\mu$eV.  Due to the openness of the system qubit in our experiment, and other possible sources of error, we expect to observe a deviation in our measured work PDF from the ideal work distribution $p^{id}(w)$. 

\begin{figure}\label{fig:wpdf_comparison}
\centering
\begin{overpic}[width=0.93\columnwidth]{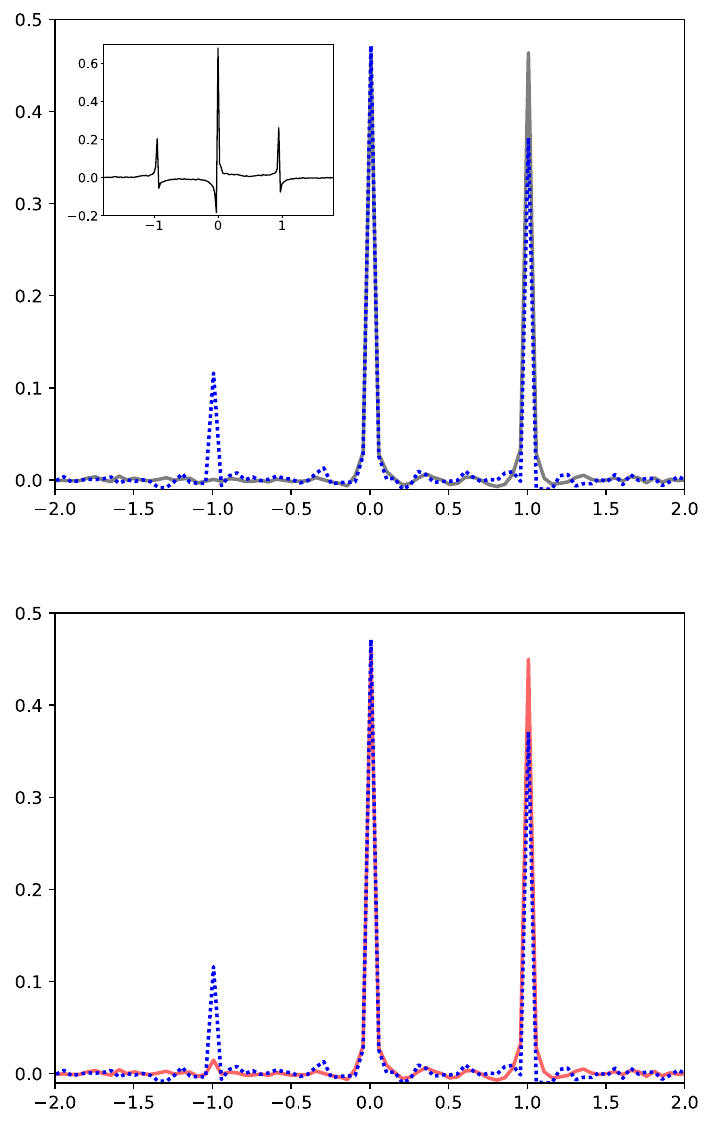}
\put(-1,100){(a)}
\put(-3,23){\rotatebox{90}{$p(W)$}}
\put(30,-2){W [$\hbar\omega$]}
\put(-1,48){(b)}
\put(-3,75){\rotatebox{90}{$p(W)$}}
\put(30,50){W [$\hbar\omega$]}
\end{overpic}
\caption{Work PDFs measured via the interferometric scheme on real quantum computer versus a quantum emulator. (a) Comparison of work PDF measured from the quantum computer (blue dotted curve) versus the work PDF computed with a noise-free quantum emulator (solid gray curve) with the qubit initialized in the pure ground state. Inset: Measurement of $2\mathbb{RE}[q_>(w)]$ on the quantum computer when the system qubit is intentionally initialized in a coherent state. (b) Comparison of work PDF measured from the quantum computer (blue dotted curve) versus the work PDF computed with a noise-free quantum emulator (solid red curve) with the qubit initialized in a thermal state at the measured effective initial temperature $T_0 = 67$ mK.}
\end{figure}
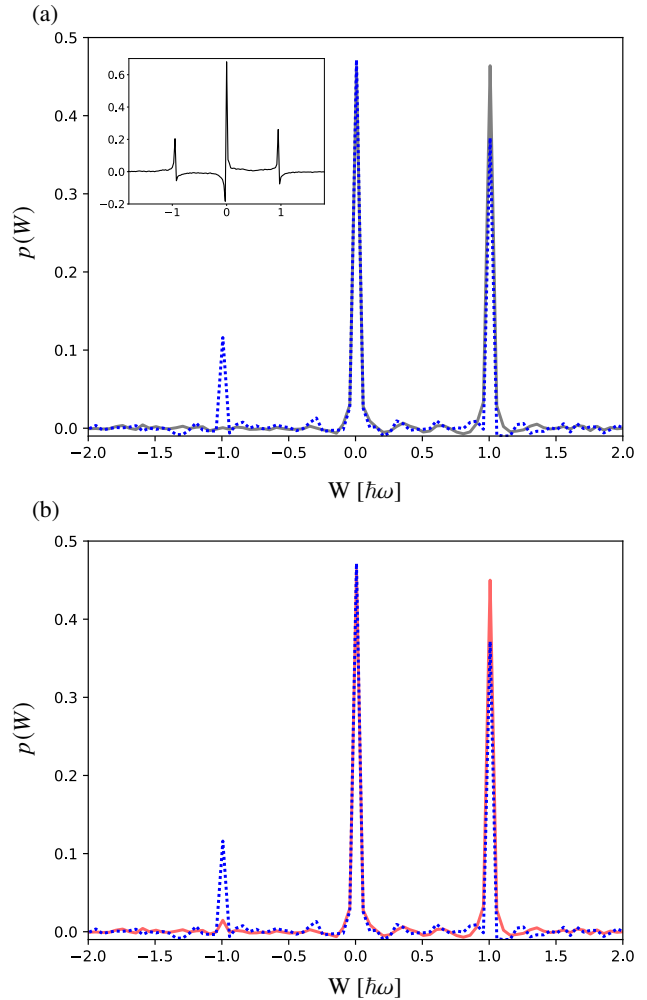

The blue dotted curve in Figure \ref{fig:wpdf_comparison}a shows twice the real part of the half-inverse-Fourier transform of the $g(u)$ (i.e., $2\mathbb{RE}[q_>(w)]$) measured on the quantum computer. For comparison, the ideal closed-system work PDF, computed from a noise-free quantum emulator (which simulates the performance of a noise-free quantum computer on a classical computer), is plotted with the solid gray curve. A number of observations are in order.

First, note that the closed-system work PDF does not exactly match the ideal work PDF $p^{id}(w)$, Eq. (\ref{eq:p-2peaks}), despite the fact that it was produced with a noise-free quantum emulator. This discrepancy is due to the finite sampling range and rate of $g(u)$. Note from Eq. (\ref{eq:q_>}) that in order to exactly recover the quantity $q_>(w)$, one needs to sample $g(u)$ on the continuum infinite set $[0,\infty)$.  Due to constraints related to access and compute time on the IBM quantum computers, we were limited to sampling 900 values for $u$. To achieve sufficient resolution and range of work values in the resulting work PDF, a step-size of $\Delta u = 0.013$ between sampled values of $u$ was chosen, resulting in a sampling range of $[0,11.5]$ for $u$ in units of $[\mu \text{eV}]^{-1}$.  The same sampling rate and range were used for all work PDF plots.

Second, we observe that the distribution derived from the quantum computer does not present any antisymmetric $1/x$-type broadening around the peaks, implying that the imaginary part of the measured quasi-probabilities was null, $\mathbb{IM}[q_{mn}]=0$. This is a clear indication that there was no coherence in the initial preparation, and we can safely consider the plotted distribution to be a genuine work PDF, i.e., $2\mathbb{RE}[q_>(w)]$ = $p(w)$.  For comparison, the inset of Figure \ref{fig:wpdf_comparison}a plots the quantity $2\mathbb{RE}[q_>(w)]$ as obtained from running the same quantum circuits on the quantum computer but intentionally initializing the $S$ qubit in the coherent state $(|0\rangle+ |1\rangle)/\sqrt{2}$. This is achieved by enacting the Hadarmard gate on the $S$ qubit at the start of the computation.  We observe that now the antisymmetric $1/x$ broadening of the peaks are clearly visible when initial coherence is present. Similar quasi-probability distributions have been studied experimentally, using the interferometric method with nitrogen vacancy centers in diamond, in Ref. \cite{Hernandez-Gomez23NPJQI9}.  Thus, the interferometric scheme allows us to confirm a lack of coherence in the initial preparation of the qubit on the quantum processor.  This reveals the utility of the scheme in characterizing sources of noise on quantum computers, such as spurious coherence in the initial state preparation of the qubits.

Third, assuming the system was genuinely open (more on this below), the absence of initial coherence means that the (unknown) initial SB state $\rho_{SB}$ was commuting with the (unknown) SB unperturbed Hamiltonian $H_{0}^{SB}$. Then, under the reasonable assumption that the initial state was factorized as $\rho_{SB}= \rho_S \otimes \rho_B$, the commutation condition $[\rho_{SB},H_{0}^{SB}]=0$ indicates that the (unknown) interaction term $H_I^{SB}$, see Eq. (\ref{eq:H0SB}) is at most a weak interaction term
\footnote{To be fully correct, the commutation condition can also hold in the special case when $H_I^{SB}$ commutes with both the system unperturbed Hamiltonian $H_0$ and the bare bath Hamiltonian $H_B$, which we exclude.}.
Already this allows us to gain a crucial information about the SB interactions. Furthermore it confirms that the initial $S$ state $\rho_S$ has no coherence relative to the bare Hamiltonian $H_0^S$.

%The work PDF obtained from the open-system qubit on the quantum computer is shown by the solid blue curve in Fig \ref{fig:wpdf_comparison}. 

%Fig \ref{fig:wpdf_comparison}a compares the open-system work PDF with one that would be obtained if the system $S$ was a closed system (black dotted curve).  The closed-system work PDF is derived from a noise-free quantum emulator, which simulates the performance of a noise-free quantum computer on a classical computer.  
%Note that the closed system PDF presents only two peaks: a peak centered at zero work and a peak centered around a positive value of work $W_+-= \hbar \omega$, equal to the energy gap of the qubit. We remerk that the S qubit has a resonant frequency of $\omega = .... GHz$, and all the other qubits on the chip have approximately (within.....) the same resonant frequency.
%The presnece of only two peaks refkects  the fact that the qubit is prepared in its ground state, so only one of two processes can occur as a consequence of the evolution generated by $U$: (i) the qubit remains in its ground state, resulting in null work, or (ii) the qubit transition to its excited state, requiring an amount of work equal to its energy gap. 

Fourth, while the simulated work distribution of the ideal, closed-system qubit presents only two peaks, in agreement with Eq. (\ref{eq:p-2peaks}), the measured work PDF of the qubit on the real quantum device (presumed to be an open system) clearly shows a third peak centered around the negative work value $W_-=-\hbar \omega$. This third peak could have two different origins.  The first origin could be that the initial state of the qubit $\rho_S$ was not the intended pure ground state. Given that there was no coherence in the initial state, this implies that it was rather in a statistical mixture of ground and excited states.  This, in turn, implies the initial state was a thermal state $\rho_S \propto e^{-\beta_0 H_0^S}$ with some effective temperature $T_0=1/(k_B \beta_0)$ ($k_B$ being Boltzmann's constant).  When there is a non-zero probability of the qubit starting in the excited state, then the driving can induce a transition from the excited state to the ground, resulting in the appearance of the third peak in the work PDF. Such a peak would emerge both in the case the system is open or closed, see Eq. (\ref{eq:p-3peaks}). 

The second origin of the third peak could be that the system $S$ was undergoing genuinely open dynamics, in which case this peak signals a transition occurring within the extended SB system.  Other sources of error, e.g., due to imperfect driving due to gate errors and measurement error, can obviously lead to imperfections in the measured characteristic function, and hence deviations from the ideal work PDF. However, such errors are not expected to result in a neat delta peak at the value of work that perfectly matches the resonance frequency of the qubits on the device (see Appendix \ref{sec:noisy_sim}).

To investigate the origin of the third peak, we examined the fidelity of the initial preparation of the $S$ qubit in the ground state.  To do so, we executed a calibration circuit comprising only a measurement of the $S$ qubit.  Across 1024 runs of the calibration circuit, the qubit was measured in the ground state with a probability of $p_0 \simeq 0.97$.  This would imply an effective initial temperature of $T_0 = 67$ mK for the qubit.
To ascertain how this initial temperature will contribute to the third peak, we performed a noise-free emulation of the circuit in Fig. \ref{fig:schematics}b with the $S$ qubit initialized in a thermal state at this temperature $T_0$. 

The resulting work PDF is plotted by the red curve in Fig \ref{fig:wpdf_comparison}b, where it can be compared with the observed work PDF from the open system qubit on the quantum computer (blue dotted curve).  We observe that, as expected, a third peak appears in the emulation of the qubit initialized in a thermal state. However, the peak is significantly smaller than the measured peak from the quantum computer. We conclude that the initial finite temperature $T_0$ of the qubit $S$ preparation is not sufficient to fully account for the intensity of the third peak. %We conclude the system was genuinely interacting with a bath during the drive and the ancilla picked accordingly transitions occurring in the SB system. 

One may object that the ancilla qubit, just like the $S$ qubit, also has an initial finite probability of being in the excited state (that is a finite initial temperature $T_0^A$) which could be responsible for the emergence of the third peak. By repeating the analysis presented in Ref. \cite{Dorner13PRL110} for an ancilla initialized in the the $|1 \rangle$ state rather than in the $|0 \rangle$ state, it is straightforward to see that this only results in an overall sign change of the expectation of its $\sigma_z$ and $\sigma_y$ Pauli matrices. Thus, when the ancilla is initially in a statistical mixture of the ground and excited states, with weights $p_1^A$, $p_0^A=1-p_1^A$, we obtain
\begin{align}
    g(u) = (1-2 p_1^A) [\langle \sigma_z^A \rangle(u) + i \langle \sigma_y^A \rangle(u) ]
\end{align}
which reduces to Eq. (\ref{eq:g=sigmas}) for $p_1^A=0$. An initial thermal state of the ancilla state, therefore, only results in an overall damping factor. To estimate the value of $p_1^A$, we again run a calibration circuit, this time solely comprising a measurement of the ancilla qubit. Across 1024 runs of the calibration circuit, the ancilla qubit was measured in the excited state with probability $p_1^A \simeq 0.01$, hence a negligible damping factor of $ 1- 2 p_1^A \simeq 0.98$. %Such a factor can always be taken care of by re-normalizing the characteristic function, by enforcing that the condition $g(0)=1$ be obeyed.

Given that the observed effective temperatures of the system and ancilla qubits cannot fully account for the third peak in the measured work PDF from the quantum computer, we therefore conclude that some other process (or processes), involving the qubit and its environment, must be responsible for this peak, confirming that the interferometric scheme is indeed measuring the work in an open quantum system.  Since the third peak is centered around a negative value of work equal to the frequency of the qubits on the chip, it is likely that the interferometric scheme is detecting an inadvertent interaction of the open system qubit with another qubit in its vicinity, i.e., cross-talk, a well-documented major source of noise on superconducting quantum computers \cite{zhao2022quantum,ketterer2023characterizing}.  

Since our system of interest is open, one could argue it might be possible to reproduce the experimental data by employing a noisy quantum emulator (a program that emulates the performance of a noisy quantum computer on a classical computer). To test this hypothesis, we executed the set of interferometric circuits on a state-of-the-art noisy quantum emulator optimized to emulate the behavior of the IBM quantum processor, but were not able to reproduce the experimental data. See Appendix \ref{sec:noisy_sim} for more details. This underscores the utility of the interferometric scheme in characterizing the noise in quantum computers, the complete details of which are clearly not fully captured by the commonly employed noise models.

\section{Towards the verification of the Jarzynski Equality}
One of the most powerful theoretical results in non-equilibrium thermodynamics is the celebrated Jarzynski equality  \cite{jarzynski1997nonequilibrium,jarzynski1997equilibrium}, which relates fluctuations in the work performed on a driven system with the free energy difference as 
\begin{align}
    \langle e^{-\beta w} \rangle = e^{-\beta \Delta F}
\end{align}
where $\beta$ is the inverse temperature at which the system is in thermal equilibrium, $w$ is the work performed in a particular drive, $\Delta F$ is the change in equilibrium free energy of the system, and $\langle \cdot \rangle$ denotes an average over independent, identical drives.  While originally derived for closed classical systems, the Jarzynski equality has been shown to remain valid as well for open quantum systems \cite{Campisi09PRL102}, under the provision that the system-bath compound is at thermal equilibrium at inverse temperature $\beta$ at the start of the driving protocol. As $\Delta F$ is the free energy difference relative to the initial and final Hamiltonians, for cyclical drivings, such as the one considered here, we have $\Delta F =0$, which simplifies the Jarzynksi equality to:
\begin{align}
    \langle e^{-\beta w} \rangle = 1.
    \label{eq:Jarz1}
\end{align}

As mentioned in the introduction, owing to the difficulty in measuring the work statistics in open quantum systems, the Jarzynski equality has yet to be experimentally verified for such systems. Our demonstration of the measurement of the work PDF for an open quantum system is therefore a crucial advance in this pursuit. Despite having achieved such a measurement, here we are not able to complete the verification as we do not have a means of measuring the temperature of the bath $B$. Given this information, we could prepare the system $S$ at this bath temperature and thereby (approximately) realize the SB thermal equilibrium initialization 
\begin{align}
    \rho_{SB} = e^{-\beta (H_0^S \otimes \mathds{1}_B+ \mathds{1}_S \otimes H^B)}/Z \simeq  e^{-\beta H_0^{SB}}/Z \label{eq:thermal}
\end{align}
where the $\simeq$ sign follows from the fact that we know that the coupling is weak.%necessary for the verification of Eq. (\ref{eq:Jarz1}). weak coupling

We are, however, able to establish whether or not our experimental data are consistent with the validity of the Jarzynski equality.  To do so, we performed a second experiment involving two sets of circuits: one set identical to the ones described above, and the other set with the the $S$ qubit intentionally prepared in its excited state, which simply requires enacting an $X$-gate on the qubit.  Just as preparation of the ground state occurred with some finite fidelity (associated to the finite effective temperature $T_0$), preparation of the excited state likewise occurs with a finite fidelity, which can be associated with a negative temperature $T_1$. Using calibration circuits consisting solely of measurement, we measured the effective initial temperature of the qubit when it was prepared in the ground state to be $T_0 = 83$ mK, and analogously measured the effective initial temperature of the qubit when it was prepared in the excited state to be $T_1 = -87$ mK.

%according to the formula
%\begin{align}
%    p_1/p_0 = e^{-\hbar \omega /T_1}
%\end{align}
%where $p_1$, $p_0$ are respectively the measured probabilities that the qubit was in the $1,0$ state, after the preparation. 
%Note that $T_1<0$ owing to the fact that with high probability the qubit is in the excited state. 
We denote with $p(w,T_1)$ the work PDF measured with the qubit initialized at an effective temperature $T_1$, and analogously, we denote with $p(w,T_0)$ the work PDF measured with the qubit initialized at temperature $T_0$. Mixing the two work PDFs via a convex linear combination with weights $r$ and $1-r$, 
\begin{align}
    p(w,T) = r p(w,T_0)+(1-r) p(w,T_1)
\end{align}
provides an estimate of what the measured work PDF would be if the system were prepared at the intermediate temperature
\begin{align}
    %\beta = (\hbar \omega)^{-1} \ln \frac{r f(\hbar \omega \beta_0)+ (1-r)f(\hbar \omega \beta_1)}{r f(-\hbar \omega \beta_10+ (1-r)f(-\hbar \omega \beta_1)}
    T = \frac{\hbar \omega}{k_B } \left[ \ln \frac{r f(\hbar \omega \beta_0)+ (1-r)f(\hbar \omega \beta_1)}{r f(-\hbar \omega \beta_0)+ (1-r)f(-\hbar \omega \beta_1)}  \right]^{-1}
    \label{eq:Tconvex}
\end{align}
where $f(x)= 1/(1+e^x)$, $\beta_0 = 1/(k_BT_0)$, and $\beta_1 = 1/(k_BT_1)$. See Appendix \ref{sec:derivation} for more details. 

Given a value of $q$ one can compute the corresponding $T$ and the associated Jarzynski integral
\begin{align}
    J(T)\doteq \int dw p(w,T) e^{-w/{(k_B T)}}\, .
    \label{eq:JT}
\end{align}
\begin{figure}\label{fig:J_vs_T}
\centering
\begin{overpic}[width=0.93\columnwidth]{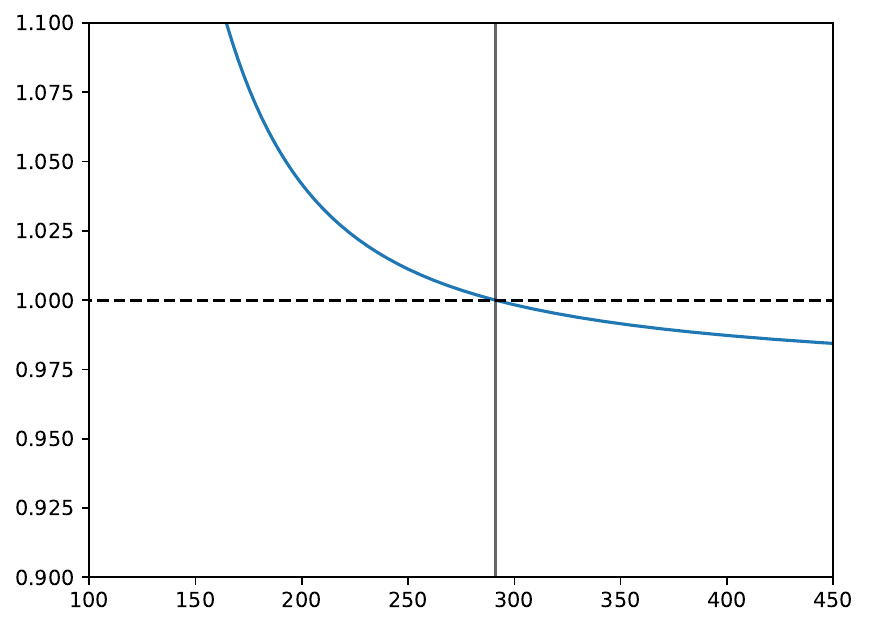}
\put(-3,37){\rotatebox{90}{$J$}}
\put(45,-2){ $T$ [mK]}
\end{overpic}
\caption{Jarsynski integral, Eq. (\ref{eq:JT}), versus the presumed temperature of the bath.}
\end{figure}
Figure \ref{fig:J_vs_T} shows the function $J(T)$. When the preparation temperature $T$ matches the unknown temperature of the surrounding bath, the overall system is (approximately) in a thermal state, Eq. (\ref{eq:thermal}), and according to theory the Jarzynski equality should be obeyed. Thus by solving the equation $J(T)=1$, for $T$ we obtain an estimate of the bath temperature 
\begin{align}
    T_B \sim 290 \text{ mK}
\end{align}
denoted by the gray vertical line in Figure \ref{fig:J_vs_T}.  This estimate appears higher than independent measurements of qubit temperatures on the IBM quantum processors reported in Ref. \cite{brand2024markovian}, ranging between $\sim 40 - 60$ mK.  However, it is a rather rough estimation due to the effects of numerous sources of error (e.g., discrete Fourier transform over truncated range, quantum computational error, shot noise, etc.). Nevertheless, we demonstrate how our method could potentially be used to estimate the temperature of qubits and their environment and propose that a bespoke experimental device with fine-tuned control over all qubit parameters, however, would likely succeed in obtaining a more accurate measurement of the temperature.

\section{Discussion}
%We have reported the measurement of the work statistics of a genuinely open quantum system. That was achieved by implementing on a quantum computer the theoretical proposal of Ref. \cite{Campisi13NJP15}, suggesting how the interferometric method of \cite{Dorner13PRL110,Mazzola13PRL110} could be used in that case. 
Leveraging the theoretical proposal of Ref. \cite{Campisi13NJP15}, based on the interferometric method of \cite{Dorner13PRL110,Mazzola13PRL110}, on a superconducting quantum computer, we have demonstrated the measurement of the work statistics of a genuinely open quantum system. While previous experiments have targeted the statistics of various stochastic quantities in open quantum systems, none have succeeded in addressing the work (i.e. total SB energy change) statistics, due to the difficulties in measuring the bath energy.  For example, Ref. \cite{Hernandez21NJP23} measures the energy change of an (open) qubit realized with a nitrogen vacancy center in diamond, but a lack of quantification of the heat prevents a measurement of the work. %Reference \cite{Smith18NJP20} reports the measurement of work of a ``pseudo'' open qubit realized with a trapped ion. In this case, the openness is mimicked by adding classical noise to the drive which induces a decoherence (unital) dynamics, rather than the presence of a thermal bath, which generally defines an open system. Accordingly, there is no heat exchange, which implies that in this special case, the change in qubit energy can be identified with work. In contrast, the interferometric scheme presented here is very general, and can be applied equally well in the case of closed system dynamics, pure decoherence dynamics, or genuinely open dynamics with arbitrary SB interaction strengths.

As the interferometric scheme detects phenomena within the extended SB system, it can be used to probe the various processes that occur in the SB system. In our specific experiments, performed on the \textit{ibm$\_$sherbrooke} quantum processor, the work measurements of the SB system evidenced that the qubit of interest was affected by cross-talk error. Cross-talk, a general term describing unintended interactions between qubits, is a predominant source of noise in current superconducting quantum computers \cite{zhao2022quantum,ketterer2023characterizing}.  Worse, since cross-talk can induce correlated and non-local errors, it poses a formidable challenge to fault-tolerant quantum computers, since quantum error correction generally relies on errors being local \cite{sarovar2020detecting}. A better understanding of cross-talk errors is therefore essential to the success of quantum computers now and in the future. We have demonstrated that the interferometric scheme can be used to detect cross-talk, potentially providing the necessary information to develop mitigation strategies against it.

It is also important to remark that the noise models that are customarily employed to simulate the dynamics of the open (i.e., noisy) quantum processor were not able to reproduce our data, thus evidencing that they cannot capture all the fine details of the SB dynamics. This implies that there is indeed room for improving such noise models and that the interferometric scheme could be useful in achieving this goal.

Last but not least, our experiments demonstrate that the interferometric scheme is a promising method for reaching the long-sought experimental verification of the Jarzynski equality for arbitrary open quantum systems \cite{Campisi09PRL102}. While we cannot claim to have achieved such a verification, as we lack a means of independent measurement of the temperature of the bath, our data are nevertheless not inconsistent with the Jarzynski equality.  This indicates that a dedicated experiment with finer controls than those available to the general user of a quantum computer could very likely reach the goal.

Using a reversed but equivalent logic, knowing that the measured work PDF of an open quantum system indeed obeys the Jarzynski equality allows one use the work PDF to estimate the temperature of the bath.  Seen from this perspective, the interferometric scheme could be used as a thermometer.  In our case, due to the error affecting the data, such an estimate of the bath temperature was very rough. Once again, with a tailor made experiment one could achieve a much higher thermometric precision.

\begin{acknowledgments}
LBO gratefully acknowledges funding from the European Union’s Horizon 2020 research and innovation program under the Marie Skłodowska-Curie grant agreement No 101063316. This research used resources of the Oak Ridge Leadership Computing Facility, which is a DOE Office of Science User Facility supported under Contract DE-AC05-00OR22725. We acknowledge the use of IBM Quantum services for this work. The authors thank Nicole Fabbri and Santiago Hern\'andez-G\'omez (CNR-INO, Florence) and Ettore Bernardi (INRIM, Turin) for the insightful discussions.

\end{acknowledgments}

\appendix
%\section{Methods}

\section{Noisy Quantum Emulation}
\label{sec:noisy_sim}
We attempt to reproduce the work PDF measurements obtained from the quantum computer of the open-system qubit using a noisy quantum emulator (a program that emulates the performance of a noisy quantum computer on a classical computer). The noisy quantum emulator is characterized by a noise model which can account for the dominant sources of error including gate errors, thermal relaxation error, and measurement error.  Gate errors are incorporated into the noise model as a depolarizing channel, with the associated probability of error given by the particular gate fidelity.  Thermal relaxation error is incorporated into the noise model as a thermal relaxation channel parameterized by the decoherence and relaxation times, $T1$ and $T2$, of the particular qubit and the wall-clock time of each gate. Measurement error, defined by the probability of measuring the qubit in the excited state given it was prepared in the ground state and vice versa, is accounted for in the post-processing of the results.

\begin{figure}
\label{fig:wpdf_noisy_sim}
\centering
\begin{overpic}[width=0.93\columnwidth]{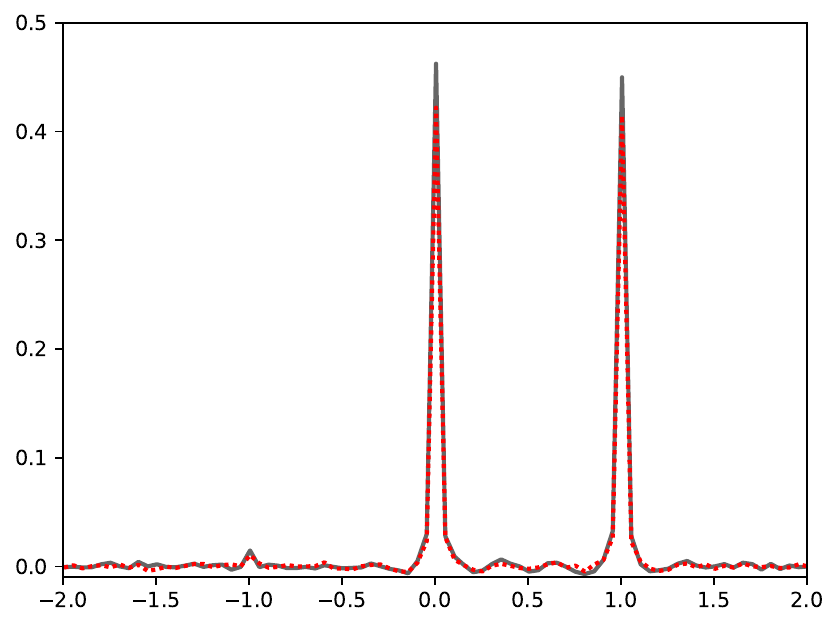}
\put(-3,37){\rotatebox{90}{$p(W)$}}
\put(45,-2){W [$\hbar\omega$]}
\end{overpic}
\caption{work pdf using a noisy versus noise-free simulator}
\end{figure}

We tailor the noisy quantum emulator to mimic the performance of the real quantum processor as closely as possible by setting the various parameters of the noise model to the values measured by calibrations performed on the qubits at the time we executed our interferometric circuits on the real quantum computer.  Namely, we use the measured gate fidelities recorded for each gate type acting on each qubit, the measured $T1$ and $T2$ decoherence times of qubits, the gate durations of the quantum processor, the recorded measurement error of the ancilla qubit, all of which are readily provided by IBM, calculated via randomized benchmarking and other standard calibration techniques. Furthermore, we initialized the $S$ qubit into a thermal state at the measured temperatures $T_0$ discussed above, to account for the initial state preparation error. We chose to initialize the ancilla qubit in its ground state, as error due to an initial thermal state of the ancilla only results in an overall damping of the peaks, which can be corrected with knowledge of the initial temperature of the ancilla.

Figure \ref{fig:wpdf_noisy_sim} compares the work PDF computed with the noisy quantum emulator (red dotted curve) to the one computed with a noise-free quantum emulator (solid gray curve), where in both cases the $S$ qubit was initialized in a thermal state at the measured effective initial temperature $T_0 = 10.7$ mK of the qubit on the quantum computer.  We observe that the peaks in the work PDF from the noisy quantum emulator are damped compared to those from the noise-free emulator.  As this damping is not compensated by the presence of new peaks, the sum of the probabilities in the work PDF from the noisy emulator no long sum to unity, in other words, we no longer observe a normalized work PDF. The overall probability contained in the estimated PDF was in fact $\simeq 0.92$. Note that the probability leakage does not originate from the ancilla initialization, which was in its ground state. Furthermore, even if one re-normalized the PDF ``by hand'', this would still not succeed in reproducing the experimental data.  We also observe that the intensity of the third peak is comparable with that from the noise-free emulator, remaining considerably lower than the experimental one. In summary, the noisy quantum emulator does not reproduce the experimental data, and worse does not even result in a bona-fide work PDF. The fact that it is unable to reproduce the results from the real quantum computer further highlights to utility of the interferometric method as a diagnostic probe of the effects of the environment on the qubits, some of which, apparently, cannot be captured with the commonly used theoretical noise models.

\section{Derivation of Eq. (\ref{eq:Tconvex})}
\label{sec:derivation}
Let $P_0(T_i),P_1(T_i)=1-P_0(T_i)$ be the populations of the two states of the S qubit in the measured preparations at temperatures $T_i$, $i=0,1$. That is:
\begin{align}
    \frac{P_1(T_i)}{P_0(T_i)}= e^{-\hbar \omega/(k_B T_i)}
    \label{eq:P1/P0's}
\end{align}
Let $R_0, R_1$ be the populations of the two states of the S qubit resulting from a convex combination of the $T_0$ and $T_1$ populations,
\begin{align}
    R_n= q P_n(T_0) + (1-q) P_n(T_1),
    \label{eq:Ralpha}
\end{align}
where $ 0 \leq q\leq 1,\, n=0,1 $.
The effective temperature $T$ of this mixture is given by the equation:
\begin{align}
     \frac{R_1}{R_0}= e^{-\hbar \omega/(k_B T)}
\end{align}
Solving for $T$ and using Eqs. (\ref{eq:P1/P0's},\ref{eq:Ralpha}) one gets Eq. (\ref{eq:Tconvex}). 

Recalling that our experiments occur in the weak-coupling regime, the SB work distribution resulting from a preparation of the $S$ qubit at temperature $\vartheta$ can be written as:
\begin{align}
    p(w,\vartheta) = \sum_{m,n,\mu,\nu} P_n(\vartheta) P^B_\nu P_{m,\mu|n,\nu} \delta (w - E_{m,\mu}+E_{n,\nu}) 
    \label{eq:pwtheta}
\end{align}
where $m,n$ label the $S$ states, $\mu, \nu$ label the environmental states, $P^B_\nu$ is the probability that the environment is initially in state $\nu$, $P_{m,\mu|n,\nu}$ is the conditional probability to transition from state $n,\nu$ of the SB system, to the state $m,\mu$ and $E_{m,\mu},E_{n,\nu}$ are the according SB eigenenergies. Due to linearity of Eq. (\ref{eq:pwtheta}) in the S populations,  $P_n(\vartheta)$, it is:
\begin{align}
     q p(w,T_0)+ (1-q) p(w,T_1)
     %= \sum_{m,n,\mu,\nu} R_n P^B_\nu P_{m,\mu|n,\nu} \delta (w - E_{m,\mu}+E_{n,\nu}) \nonumber\\
     = p(w,T)
\end{align}
that is, by mixing the two work PDF's $p(w,T_0),p(w,T_1)$ resulting from the $S$ qubit being prepared at $T_0,T_1$, respectively, with wiegths $q$, $1-q$, one gets the work PDF that would result from a preparation at temperature $T$ given by Eq.  (\ref{eq:Tconvex}).

%In hindsight that is not too surprising. Imagine, for simplicity the case when the noise only affects the S qubit gate $U$, which would accordingly result in the unitary evolution $U \rho U^\dagger$, to be replaced by a quantum channel 
%$\sum_k A_k \rho A_k^\dagger$ with Kraus operators $A_k$. This situation is exactly the same as that studied in Ref. \cite{Goold14PRE90} (the only difference being that the roles of system and bath are inverted). According to the results in Ref. \cite{Goold14PRE90}, then the interferometric method would give the pdf of change of the energy $\Delta E$ of the system S. Note that this would be a properly normalised pdf, but it would not be the work pdf. Now if noise is added to the rest of the gates involving the S qubit alone, those result in replacing the phase shift operators $e^{-H_S u/\hbar}$ with non-unitary channels, which in turn would spoil the interferometric scheme, with the consequent normalisation issues.
%Adding some low level noise to the ancilla as per our emulator runs, would only spoil the S energy change pdf further. 

\bibliography{references}

\end{document}